\documentclass[useAMS,usenatbib]{mnras}
\usepackage{times}
\usepackage{color}
\usepackage{breakurl}
\usepackage{color}
\usepackage{mathrsfs}
\usepackage{multicol}        
\usepackage{bm}		
\usepackage{pdflscape}	

\usepackage{amsfonts,amsmath,amssymb,mathrsfs}
\usepackage{graphicx,epsf,epsfig}
\usepackage{inputenc}
\usepackage{subfigure}
\usepackage{bm}
\usepackage{longtable}
\usepackage[usenames,dvipsnames]{xcolor}
\usepackage{multirow}

\usepackage{hyperref}
\usepackage{times}
\usepackage{color}

\usepackage{appendix}

\def\be{\begin{equation}}
\def\ee{\end{equation}}
\def\bea{\begin{eqnarray}}
\def\eea{\end{eqnarray}}

\title[Accretion disk luminosity for black holes surrounded by dark matter]{
Accretion disk luminosity for black holes surrounded by dark matter}

\author[K.~Boshkayev, A.~Idrissov, O.~Luongo and D.~Malafarina]{Kuantay Boshkayev,$^{1,2,3}$\thanks{kuantay.boshkayev@nu.edu.kz, kuantay@mail.ru} Anuar Idrissov,$^4$\thanks{anuar.idrissov@nu.edu.kz} Orlando Luongo$^{5}$\thanks{orlando.luongo@lnf.infn.it} and Daniele Malafarina$^4$\thanks{daniele.malafarina@nu.edu.kz} \\
$^1$National Nanotechnology Laboratory of Open Type, Department of Theoretical and Nuclear Physics, Al-Farabi Kazakh National University,\\ Al-Farabi ave. 71, 050040 Almaty , Kazakhstan.\\
$^2$Energetic Cosmos Laboratory, Nazarbayev University, 53 Kabanbay Batyr, 010000 Nur-Sultan, Kazakhstan.\\
$^3$Fesenkov Astrophysical Institute, Observatory 23, 050020 Almaty, Kazakhstan\\
$^4$Department of Physics, Nazarbayev University, 53 Kabanbay Batyr, 010000 Nur-Sultan, Kazakhstan.\\
$^5$Istituto Nazionale di Fisica Nucleare, Laboratori Nazionali di Frascati, 00044 Frascati, Italy.
}

\begin{document}

\date{\today}
\maketitle

\begin{abstract}
We consider the observational properties of a static black hole space-time immersed in a dark matter envelope. We thus investigate how the modifications to  geometry, induced by the presence of dark matter affect the luminosity of the black hole's accretion disk. We show that the same disk's luminosity produced by a black hole in vacuum may be produced by a smaller black hole if surrounded by dark matter under certain conditions. In particular, we demonstrate that the luminosity of the disk is  markedly altered by dark matter's presence, suggesting that mass estimation of distant super-massive black holes may be changed if they are immersed in dark matter. We argue that a similar effect holds in more realistic scenarios and we discuss about the refractive index related to dark matter lensing. Hence we show how this may help explain the observed luminosity of super-massive black holes in the early universe.
\end{abstract}




\section{Introduction}\label{sec:1}

The recent observation of the 'shadow' of the super-massive black hole candidate in the galaxy M87 \citep{2019ApJ...875L...1E,2019ApJ...875L...6E} proves that extreme compact objects reside at the center of galaxies. Such objects are not only seen in the electromagnetic spectrum, being sirens for
gravitational waves and \emph{de facto} bringing about the novel approach of observing our Universe based on
\emph{multi-messenger
astronomy} \citep{2019NatRP...1..585M}.

However the current theoretical models for the formation of super-massive black holes do not explain the distribution of black hole masses with distance \citep{2013ARA&A..51..511K}. It is not clear how and when super-massive black holes formed, i.e. their formation is far less understood than that of their light, stellar mass, counterparts. We only know a large number of super-massive black hole candidates exceed one billion Solar masses and they have been observed in the early universe \citep{nojo}. The most striking example is the black hole candidate ULAS J1342+0928, with a mass measured in the range of $\sim 800\cdot 10^6 M_{\bigodot}$ and located at $z=7.54$, i.e. quite close to the Big Bang \citep{bagna}. How did such enormous objects form in such a short time?

Currently, with the only exceptions of Sgr-A* in the Milky Way and the super-massive black hole candidate in the galaxy M87, the most widely adopted method used to determine the mass of such objects consists in measuring the spectra coming from their accretion disks \citep{abramo}. For example, black holes accretion at super-critical values does not imply that it radiates at super-Eddington luminosity \citep{shapirobook}. This fact is due to the properties of the accretion disk.
Several assumptions are needed in order to produce the accretion disk models, including, of course, the assumption regarding the geometry in which the accretion disk exists \citep{2013LRR....16....1A}.

The theory of black hole accretion was developed by \cite{novikov1973, page1974} and it has been successfully applied to astrophysical black hole candidates to explain the features of their observed spectrum for many years. However observations are almost always interpreted assuming the presence of a black hole in vacuum (i.e. the Kerr metric). Only in recent years several attempts have been made at studying the theoretical properties of accretion disks in a geometry that departs from the Kerr line element \citep{2009CQGra..26u5006H, 2009PhRvD..79f4001H, 2011ApJ...731..121B,2013PhRvD..88f4022B}.
At the same time we know that galaxies are surrounded by dark matter halos which are responsible for the observed rotation curves of stars far away from the center \citep{dark1,dark2,dark3}.

Dark matter halos extend throughout galaxies from the outer regions to the center, and it is reasonable to consider the possibility that the central distribution may affect the geometry of the region. Some models have been developed to describe dark matter distributions at galactic centers \citep{dark4} and dark matter profiles around black holes \citep{konoplya2019}. It is reasonable to suppose that such dark matter profiles will have non-negligible relativistic effects close to the galaxies' cores and therefore modify the properties of the light emitted by accretion disks\footnote{In addition, the possibility that dark matter is responsible for the observed cosmic speed up has been recently investigated, see  \citep{luongo2018} for details.}.

In this paper we aim at studying this effect qualitatively by constructing a simple toy model.
For clarity, we will assume a static, spherically symmetric super-massive black hole surrounded by dark matter. Our approach to the study of the spectrum of the disk is then based on a simple model for the dark matter envelope. To do so, we shall assume that the interaction cross section of dark matter with normal matter vanishes, i.e. the dark matter envelope does not interact with the normal matter in the disk, thus allowing to model the trajectories of baryonic matter in the accretion disk as moving on circular geodesics.
We will also assume that the dark matter envelope extends inside the boundary radius of stable circular orbits in the accretion disk. This is reasonable if, for example, the dark matter content is made of weakly interacting particles travelling on various orbital trajectories.
We then evaluate the motion of test particles in the accretion disk in the geometry produced by the black hole plus dark matter source and compute the effects of the presence of dark matter on the spectrum of the accretion disk, under simple reasonable assumptions for the emission profile. We show that, under certain circumstances, the  flux and luminosity of the disk is noticeably increased by the presence of dark matter, especially in the inner parts of the accretion disk.
Furthermore, the presence of the dark matter envelope alters the spectral luminosity of the disk at all
frequencies, with respect to the case of a black hole in vacuum. This feature could in principle be tested against observations, once a more realistic version of the model and sufficiently accurate observational data become available. In this respect, we investigate the refractive index and we show its shape, comparing the cases with and without dark matter. The discrepancy between these cases is also discussed and interpreted.
These results suggest the possibility that super-massive black holes in the distant universe may be less massive than previously estimated. Further, this would highlight
the role of dark matter at galactic cores, as it acts as a source for increasing or decreasing the accretion disk's luminosity.

The paper is organized as follows. The formulation of the model and methodology are described in section~\ref{sec:modelandmet}. The  theoretical setups and main results, together with a detailed discussion, are presented in section~\ref{sec:results}. Finally,  conclusions and perspectives are summarized in section~\ref{sec:concl}.


\section{Model and methodology}\label{sec:modelandmet}

Recently a toy model for a dark matter cloud located at the galactic center without the presence of any super-massive black hole was considered by some of us \citep{bosh2019}. By analyzing the motion of test particles in the central regions of the galaxy we showed that at distances larger than 100 astronomical units there was no observable difference between the gravitational fields of a black hole and a dark matter core of the same total mass. However, black holes are likely to exist at the center of galaxies, as suggested by emission lines of quasars \citep{davidis}.
In this work we consider a black hole located at the center of a dark matter distribution and estimate the radiative flux emitted by the accretion disk and its spectral luminosity distribution as observed at infinity. To this end, we must make a choice for the density profile of the dark matter envelope. Since our argument is mostly qualitative, we shall consider a simple exponential profile, that will suffice to approximate a reasonable matter distribution.

Thus the following density profile has been adopted for the dark matter distribution
\begin{equation}\label{eq:den}
\rho(r) = \rho_0  e^{-\frac{r}{r_0}}, \quad r\geq r_b,
\end{equation}
where $\rho_0$ is the dark matter density at $r=0$, $r_0$ is the scale radius and $r_b$ is the inner edge of the dark matter envelope, i.e. the boundary separating the interior vacuum region, described by the gravitational field of a black hole, and the exterior dark matter distribution. In principle the inner edge of the dark matter envelope could be placed at any radius, though it is reasonable to assume that $r_b$ should be greater than the black hole's event horizon.
Since in our model the dark matter envelope never reaches the limit $r=0$, the parameter $\rho_0$ must be considered here to be a characteristic density.
The corresponding dark matter mass profile is obtained as
\begin{eqnarray}\label{eq:mdmprof}
M_{DM}(r)&=&\int_{r_b}^r 4 \pi  \tilde{r}^2 \rho (\tilde{r}) \, d\tilde{r} \\ \nonumber
&=&8 \pi r_{0}^3 \rho_{0}\Big[e^{-x_b} \left(1+x_b+\frac{x_b^2}{2}\right)  \\ \nonumber
&&- e^{-x}\left(1+x+\frac{x^2}{2}\right)\Big], \quad r>r_b
\end{eqnarray}
where $x=r/r_0$ and $x_b=r_b/r_0$. For vanishing $r_b$ Eq.~\eqref{eq:mdmprof} reduces to the one obtained by \cite{sofue2013}.
\begin{figure}
\centering
\includegraphics[width=1\columnwidth,clip]{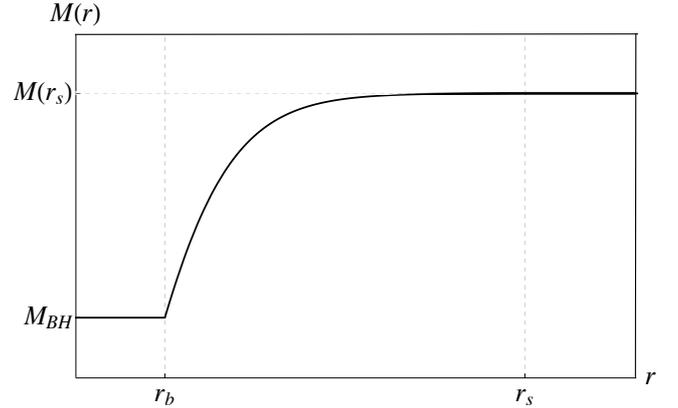}
\caption{Schematic illustration of the total mass profile.}\label{fig:massprof}
\end{figure}
The total mass profile of the system black hole plus dark matter is then defined as follows
\begin{equation}\label{eq:massprof}
 M(r)=M_{BH}+M_{DM}(r)
\end{equation}
where $M_{BH}$ is the black hole mass which is a free parameter of the model.  More explicitly Eq.~\eqref{eq:massprof} can be written as
\begin{equation}\label{eq:massprof2}
    M(r)=\left\{
                \begin{array}{lll}
                  M_{BH},  \quad \qquad \qquad \qquad r_g < r \leq r_b ,\\
                  M_{BH}+M_{DM}(r), \quad \quad r_b \leq r \leq r_s ,\\
                 M_{BH}+M_{DM}(r_s), \, \, \, \quad r_s \leq r ,
                \end{array}
              \right.
\end{equation}
where $r_g=2 M_{BH}$ is the gravitational radius of a black hole. Hereafter we use the geometric units setting $G=c=1$.
In Fig.~\ref{fig:massprof} the schematic illustration of the total mass is shown, where $r_s$ is the surface radius of the dark matter envelope, i.e. the outer edge of dark matter profile, as obtained by the solution of the Tolman-Oppenheimer-Volkoff (TOV) equation.


\subsection{The Tolman-Oppenheimer-Volkoff equation}

To analyze the properties of the system composed of a black hole and dark matter envelope, the line element is chosen in the standard static and spherically symmetric form as
\begin{equation}\label{eq:le}
d s^2=e^{N(r)} d t^2 - e^{\Lambda(r)} d r^2 - r^2 \left(d \theta^2 + \sin^2 \theta d\varphi^2\right),
\end{equation}
where $(t,r,\theta,\varphi)$ are the time and spherical coordinates, $N(r) $ and $\Lambda(r)$ are the sought metric functions.

In order to get a stable (static) spherical layer (envelope) of dark matter around the black hole one has to
match the inner black hole space-time with the outer, matter filled, solution describing dark matter at the boundary $r_b$. This can be done following the same procedure to find interior solutions for the Schwarzschild space-time, since, as it is well known, the matching conditions to be satisfied are the same \citep{1996PhRvD..54.4862F}.
Hence we need to solve the TOV equations, which read
\begin{eqnarray}
\frac{d P(r)}{d r}&=&-(\rho(r) +P(r)) \frac{M(r)+4 \pi r^3 P(r)}{r(r-2 M(r))} ,\\
\frac{d N(r)}{dr}&=&-\frac{2}{\rho(r)+P(r)}\frac{d P(r)}{d r} ,
\end{eqnarray}
where $P(r)$ is the dark matter pressure, $\rho(r)$ is given by Eq.~\eqref{eq:den} and $M(r)$ is determined by Eq.~\eqref{eq:massprof2}.
At the boundary between interior vacuum and the exterior dark matter the following conditions are defined for $\rho(r_b), P(r_b)$ and $N(r_b)$:
\begin{eqnarray}
\rho(r_b)&=&\rho_b=\rho_0 \, e^{-\frac{r_b}{r_0}}, \\
P(r_b)&=&P_b, \\
N(r_b)&=&N_b=\ln{\left(1-\frac{r_g}{r_b}\right)},
\end{eqnarray}
where $\rho_0$, $r_0$ and $r_b$ are the free parameters of the model, $P_b$ is chosen to be equal to its Newtonian limit as a test value (see appendix \ref{app:A}). The TOV equation must be integrated till the pressure vanishes $P(r)=0$. This condition yields the radius of the dark matter envelop surface $r=r_s$, so that the observed total mass of the system is equal to $M(r_s)$.

Correspondingly, $N(r)$ and $\Lambda(r)$ are given as
\begin{equation}\label{eq:bc}
    e^{N (r)}=\left\{
                \begin{array}{lll}
                  1-\frac{r_g}{r}, \quad \qquad r_g < r \leq r_b ,\\
                  e^{N_{r} (r)}, \, \qquad \quad r_b \leq r \leq r_s ,\\
                   1-\frac{2 M(r_s)}{r}, \, \, \quad r_s \leq r ,
                \end{array}
              \right.
\end{equation}
 and
\begin{equation}
    e^{\Lambda (r)}=\left\{
                \begin{array}{lll}
                  \left(1-\frac{r_g}{r}\right)^{-1},\qquad \quad r_g < r \leq r_b ,\\
                  \left(1-\frac{2 M(r)}{r}\right)^{-1}, \, \quad r_b \leq r \leq r_{s} ,\\
                  \left(1-\frac{2 M(r_s)}{r}\right)^{-1}, \, \, \, \, \, r_s \leq r .
                \end{array}
              \right.
\end{equation}
where $N_{r} (r)$ is given by the numerical solution of the TOV equation.

Notice that, since the density of dark matter is not zero at $r=r_b$, the matching is not continuous as the first derivatives of the metric have a jump at the boundary. This can be interpreted, following the usual procedure, with the presence of a massive surface layer at $r_b$ \citep{1966NCimB..44....1I,1967NCimB..48..463I}.


\subsection{Radiative flux and spectral luminosity}

To model the accretion disk, we follow the standard treatment and assume that particles follow circular geodesics in the equatorial plane ($\theta=\pi/2$). Then particles in the disk will have specific energy, angular momentum and angular velocity depending on their distance from the central object.

We can then study the characteristics of the radiative flux ${\cal F}$ (i.e. the energy radiated per unit area per unit time) emitted by the accretion disk from the following formula
\begin{equation}\label{eq:flux}
    {\cal F} (r)=- \frac{\dot{m}}{4\pi \sqrt{g}} \frac{\Omega_{,r}}{(E-\Omega L)^2} \int_{r_i}^{r} (E-\Omega L) L_{,\tilde{r}} d \tilde{r} ,
\end{equation}
where $\dot{m}$ is the mass accretion rate, assumed to be constant, $g$ is the determinant of the metric tensor of the three-sub-space $(t, r, \varphi)$ i.e. $\sqrt{g}=\sqrt{g_{tt}g_{rr}g_{\varphi\varphi}}$, $\Omega$ is the orbital angular velocity, $E$ is the energy per unit mass and $L$ is the orbital angular momentum per unit mass of test particles in the disk. Also, $r_i$ is the radius of the innermost stable circular orbit of the disk, which is defined via the condition $dL/dr=0$. These quantities, evaluated on the equatorial plane $\theta=\pi/2$ in a spherically symmetric, static space-time, take the form
\begin{eqnarray}
\Omega(r)&=& \frac{d \varphi}{dt}=\sqrt{- \frac{\partial_{r} g_{tt}}{\partial_r g_{\varphi\varphi}}}, \\
E(r)&=&u_{t}=u^{t}g_{tt}, \\
L(r)&=&-u_{\varphi}=-u^{\varphi}g_{\varphi\varphi} =-\Omega u^{t}g_{\varphi\varphi}, \\
u^{t}(r)&=&\dot{t}=\frac{1}{\sqrt{g_{tt} + \Omega^{2} g_{\varphi\varphi}}},
\end{eqnarray}
where $\partial_r$ is the derivative with respect to the radial coordinate,  dot is the derivative with respect to the proper time, $u_{t}$ and  $u_{\varphi}$ are the co-variant time and angular components of the four velocity (for details see \cite{joshi2011}).

For better presentation of our results it is convenient to introduce new dimensionless functions by defining $\Omega^{*}(r)=M_{T}\Omega(r)$, $L^{*}(r)=L(r)/M_{T}$ where $M_{T}=M(r_s)$ for the system with a black hole and dark matter and $M_{T}=M_{BH}$ for the system without dark matter. Then in our dimensionless units the energy remains unchanged, i.e. $E^{*}(r)=E(r)$.

Since the flux is not directly observable, the luminosity that reaches an observer at infinity ${\cal L}_\infty$ (energy per unit time) is a more useful quantity to consider (see \citet{joshi2014}). The differential of the luminosity ${\cal L}_{\infty}$
can be estimated from the flux ${\cal F}$ by the following relation of \cite{novikov1973, page1974}
\begin{equation}\label{eq:difflum}
 \frac{d{\cal L}_{\infty}}{d\ln{r}}=4\pi r \sqrt{g}E{\cal F}(r).
\end{equation}
%



Another characteristics of the accretion disk, the spectral luminosity distribution observed at infinity is given by
\begin{equation}\label{eq:speclum}
    \nu{\cal L_{\nu,\infty}}=\frac{15}{\pi^4}\int_{r_i}^{\infty}\left(\frac{d{\cal L_{\infty}}}{d \ln{r}}\right)\frac{(u^t y)^4/{\cal F}^{*}}{\exp[u^t y/{\cal F}^{{*}{1/4}})]-1} d\ln {r},
\end{equation}
where  $y=h\nu/kT_{*}$, $h$ is the Planck constant, $\nu$ is the frequency of the emitted radiation, $k$ is the Boltzmann constant, $T_{*}$ is the characteristic temperature, $u^t$ is expressed in terms of the characteristic redshift $z$ as $u^t(r)=1+z(r)$ (for details see \cite{joshi2014}) and ${\cal F}^{*}(r)=M_{T}^2{\cal F}(r)$.
The above expression provides the luminosity measured by distant observers as a function of the frequency under the  assumption that the gas in the accretion disk emits a blackbody spectrum. Taking into account Eq.~\eqref{eq:difflum}, then  Eq.~\eqref{eq:speclum} can be written in the form

\begin{equation}
\nu{\cal L_{\nu,\infty}}=\frac{60}{\pi^3}\int_{r_i}^{\infty}\frac{\sqrt{g} E }{M_T^2}\frac{(u^t y)^4}{\exp[u^t y/{\cal F}^{{*}{1/4}})]-1} dr.
\end{equation}
The last relation is conveniently dimensionless since it is normalized with respect to mass $M_T$.

As it is well known, astrophysical observations of black hole candidates can not rely directly on measurements of the spectral luminosity. Typically emission lines from the central regions of the accretion disk are used to probe the strong gravity regime and obtain information on the object's properties such as mass and angular momentum. Such emission lines are in the X-ray spectrum of frequencies and the most prominent and most widely used is the $K\alpha$ line of Iron at $6.4$ keV (see for example \cite{reynolds1999}).
The broadening of the $K\alpha$ Iron line due to gravitational redshift contains information about the gravitational potential, while the Doppler shift of the line contains information about the accretion disk's rotation and as a consequence on the source's angular momentum
(see \cite{bambi2017} for a review).
Therefore the $K\alpha$ line of Iron is an excellent tool to probe experimentally the geometry in the vicinity of black hole candidates (see for example \cite{bambi2013}).
It is reasonable to suppose that the presence of a dark matter envelope will affect the $K\alpha$ Iron line. In particular a matter distribution in the vicinity of the black hole will alter the broadening of the line thus introducing some uncertainties in the measurement of the black hole's mass.

Using the above formalism we can now analyze the effects that adding a dark matter envelope has on the light emitted by the gas in the accretion disk surrounding a supermassive black hole candidate.


\section{Theoretical setups and results}\label{sec:results}

We here discuss the numerical and theoretical outcomes of the model outlined above, in the context of super-massive black holes. We start by setting the model's free parameters as: $M_{BH}=5\times10^8 M_\odot=4.933$ astronomical units (AU), $r_b=11M_{BH}/2=27.133$AU, $r_0=10$AU and $\rho_0$ varying from $0.75\times10^{-5}$ AU$^{-2}$ to $3\times10^{-5}$ AU$^{-2}$, see\footnote{Note, for density 1AU$^{-2}\approx60.173$ g/cm$^3\approx8.9\times10^{23}M_{\odot}$/pc$^3$. } Figs.~\ref{fig:O_DM_Kerr}-\ref{fig:spec_Lum_DM_Kerr}.  In Table~\ref{tab:example} we present initial inputs: the characteristic density $\rho_0$ and the corresponding pressure $P_b$ at $r=r_b$, and the results of the numerical integration: the radius of the innermost stable circular orbits $r_i$, the surface radius of the dark matter envelope $r_s$ and the mass of dark matter $M_{DM}(r_s)$ .

\begin{table}
 \caption{Physical parameters of the dark matter envelope. The black hole case is shown without dark matter in the second line, for which $r_i=6M_{BH}$, where $M_{BH}=74/15\approx 4.933 AU$. In the case with dark matter, the choice of the parameters $\rho_0$ and $P_b$ determines the innermost stable circular orbit radius $r_i$, the thickness  of the dark matter envelope $r_s$ and the total mass of the dark matter envelope $M_{DM}(r_s)$ in units of $M_{BH}$} 
 \label{tab:example}
 \begin{tabular}{ccccc}
  \hline
  $\rho_0$ & $P_b$ & $r_i$ & $r_s$ & $M_{DM}(r_s)$ \\
   ($ AU^{-2}$) & ($ AU^{-2}$) & ($AU$) & ($AU$) &($ M_{BH}$)\\
  \hline
  0 & 0 & 29.600 & - & 0 \\

  $0.75\cdot10^{-5} $ & $3.132\cdot10^{-8}$ & 29.047 & 220.885 & $1.873\cdot10^{-2}$\\

  $1.50\cdot10^{-5}$ & $6.313\cdot10^{-8}$ & 28.483 & 227.407 & $3.747\cdot10^{-2} $\\

  $2.25\cdot10^{-5}$ & $9.545\cdot10^{-8}$ & 27.910 & 228.088 & $5.621\cdot10^{-2}$ \\

  $3.00\cdot10^{-5}$ & $12.828\cdot10^{-8}$& 27.331 & 228.790 & $7.494\cdot10^{-2}$\\
  \hline
 \end{tabular}
\end{table}

The TOV equations have been solved for selected values of $\rho_0$ from $r_b$ outwards, having set the initial value of the density as $\rho_b=\rho_0\exp{(-r_b/r_0)}$ and the corresponding value of the pressure $P_b$ (see appendix \ref{app:A}). These terms  start from the boundary $r_b$, i.e. where the dark matter distribution begins, until the distance $r_s$ at which the pressure vanishes, $P(r_s)=0$.
The value of $P(r_b)$ chosen for the integration of the TOV equation is obtained from the Newtonian limit. However, while in the Newtonian case the envelope extends until infinity, the relativistic corrections ensure that the relativistic pressure profile $P(r)$ vanishes at a finite radius $r_s$.

The solution of the TOV equations produces a stable thin layer of dark matter. By increasing the pressure $P_b$ by a factor of $\sim 1.5 $, for a fixed $\rho_0$, the thickness (width) of the dark matter layer increases,  i.e $r_s$ increases. Once $r_s$ is known, the metric functions, particularly $N(r)$ must be fixed to satisfy the boundary conditions Eq.~\eqref{eq:bc}.

We  notice that the solution of the TOV equation is numerical for both pressure and metric function $N(r)$. If the condition $P(r=r_s)=0$ yields $r_s$ to be the surface radius of the dark matter envelope, then the metric functions $N$ and $\Lambda$ must fulfil the condition on the surface $N(r_s)=-\Lambda(r_s)$. In practice the numerical value of $N_{n}(r_s)$ from the solution of the TOV equation is not equal to $-\Lambda(r_s)$. To satisfy the boundary conditions, i.e.
\begin{equation}\label{eq:bc2}
    e^{N (r)}=\left\{
                \begin{array}{ll}
                  1-\frac{r_g}{r_b}, \quad \qquad r = r_b ,\\
                  1-\frac{2 M(r_s)}{r_s}, \, \, \quad r = r_s ,
                \end{array}
              \right.
\end{equation}
the function $N_{n}$ must be redefined. Clearly, at $r=r_b$ the numerical function $N_{n}$ is suitable, albeit it is not well defined at $r=r_s$.
Thus, the suitable function turns out to be redefined by
\begin{equation}
N_{r}(r)=N_{n}(r)-\left[N_{n}(r_s)-\ln{\left(1-\frac{2M(r_s)}{r_s}\right)}\right]\frac{r-r_b}{r_s-r_b}, \qquad
\end{equation}
which fully satisfies the boundary conditions Eq.~\eqref{eq:bc2}.

Once the metric functions are evaluated the value of the radius of the innermost stable circular orbits, $r_i$ can be obtained following the standard procedure from the condition $dL/dr=0$, that holds at $r_i$ for a test particle on circular geodesic.

\subsection{Numerical results}

From the numerical results we have three possibilities for the location of the DM envelope and the innermost stable circular orbit:
\begin{itemize}
    \item[(i)] If the whole dark matter envelope is inside $6M_{BH}$. i.e. $r_b<r_s<6M_{BH}$, then $r_i>6M_{BH}$.
    \item[(ii)] If the whole dark matter envelope is outside $6M_{BH}$, i.e. $r_b>6M_{BH}$, then $r_i=6M_{BH}$ as in the vacuum case.
    \item[(iii)] If a part of dark matter is inside $6M_{BH}$ and another part is outside $6M_{BH}$, i.e. $r_b<6M_{BH}<r_s$, then $r_b<r_i<6M_{BH}$.
\end{itemize}
This result is similar to the one obtained by \cite{konoplya2019}, where the photon sphere radius has been calculated for a black hole surrounded by dark matter. In our computations we focus on the third case where $r_b<r_i<6M_{BH}$,
since case (i) has small physical significance, while case (ii) does not depart significantly from the case of a black hole in vacuum.
In particular in the limiting case of a DM envelop extending from $r_b=r_i=6M_{BH}$ outwards, all the DM particles are located outside the Schwarzschild ISCO and there will be no considerable differences in the disk's flux and luminosity with respect to the Schwarzschild case. This is due to three factors, namely (a) the geometry for $r<r_i$ which is given by the Schwarzschild metric, (b) the fact that the DM particles do not interact with the gas in the accretion disk and (c) the low mass of the DM envelope as compared to the black hole mass.

\begin{figure*}
\centering
\subfigure[Angular velocity versus radial distance in the static space-time \label{fig:O_DM}]{\includegraphics[width=3.2in]{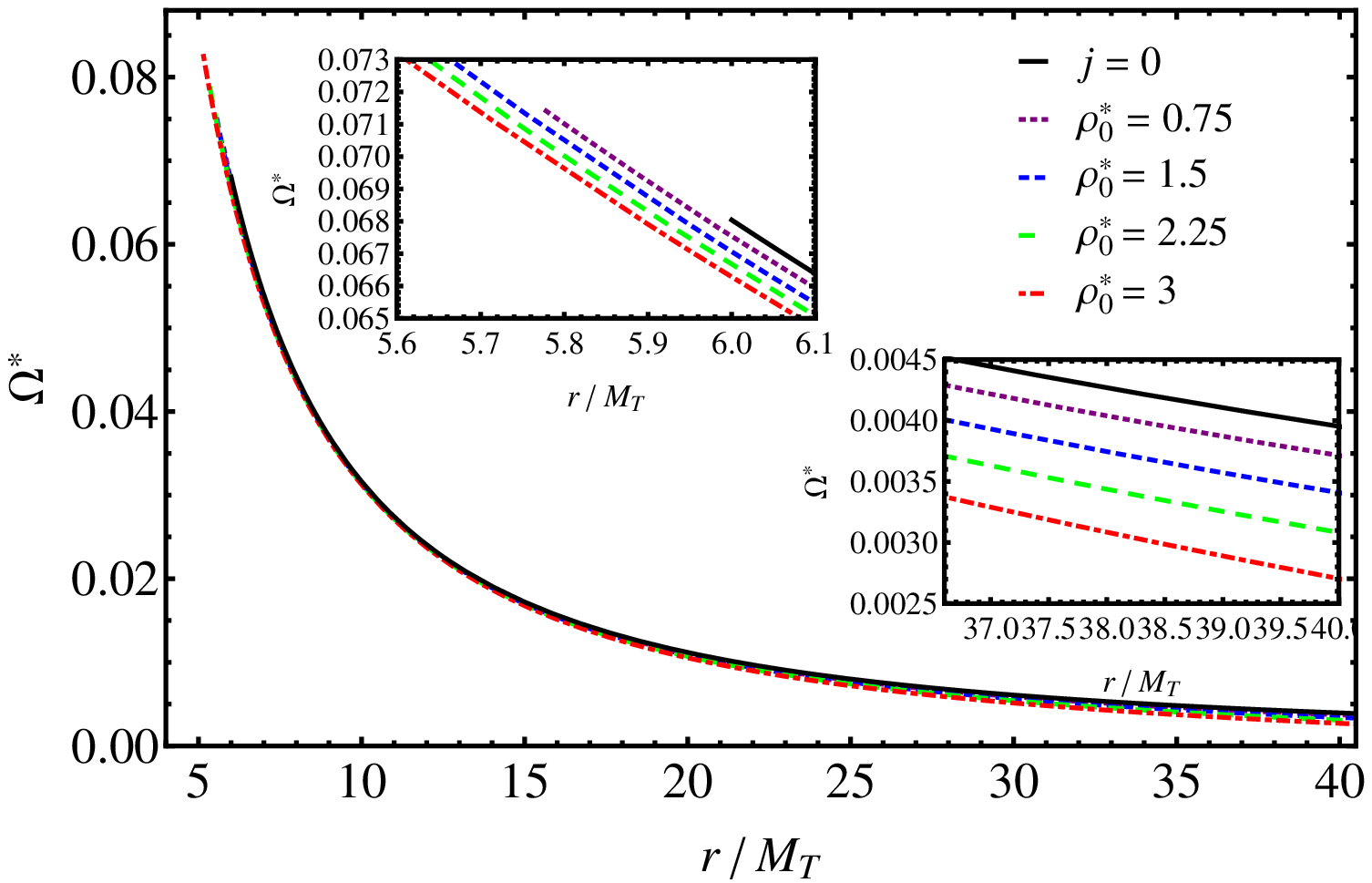}}
\subfigure[Angular velocity versus radial distance in the Kerr space-time\label{fig:O_Kerr}]{\includegraphics[width=3.2in]{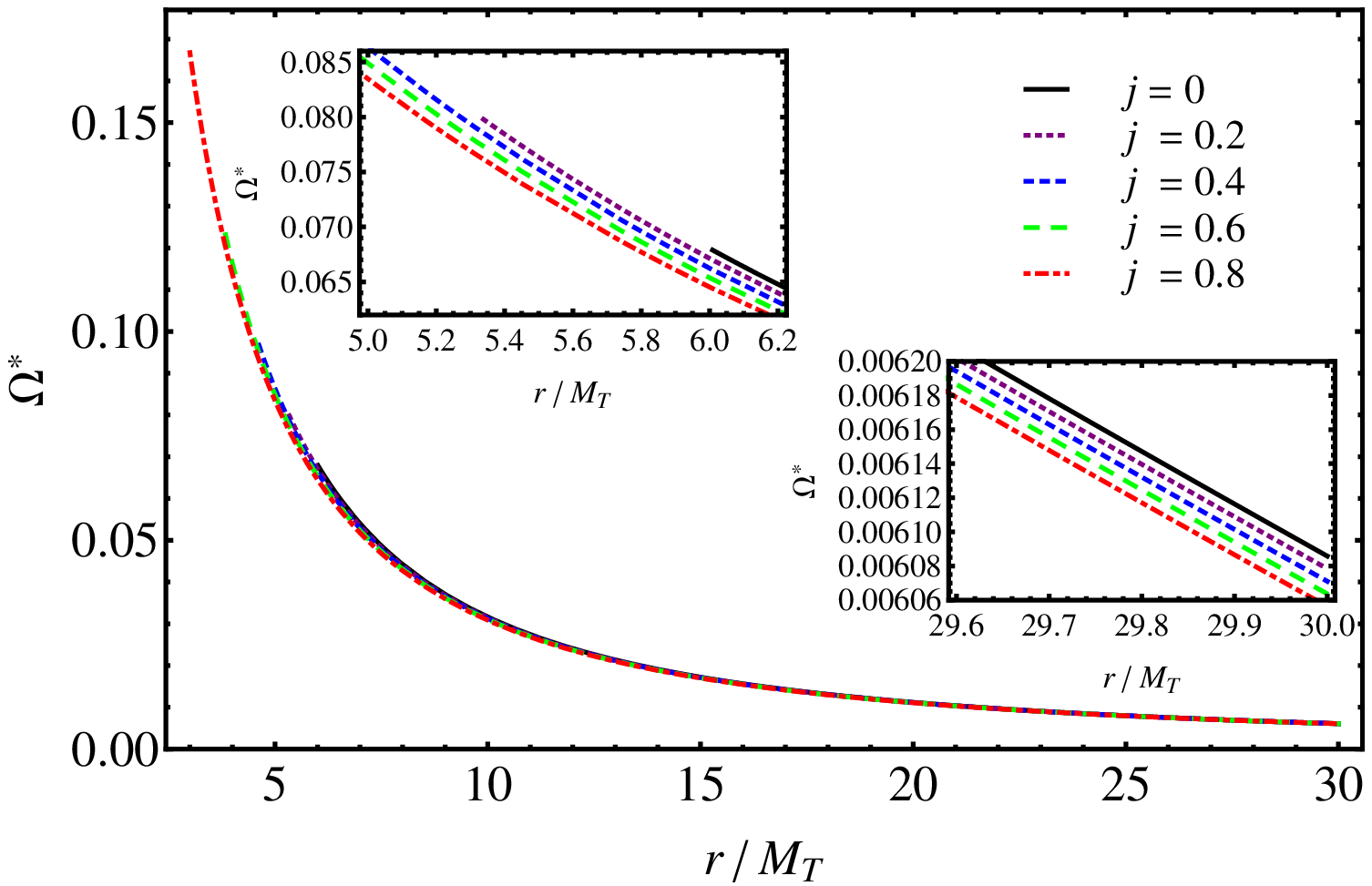}}
\caption{Numerical evaluation of the orbital angular velocity $\Omega^*$ of test particles in the presence of dark matter for a static space-time \ref{fig:O_DM} and for the Kerr black hole in the absence of dark matter\ref{fig:O_Kerr} as functions of $r/M_T$. In both figures the solid curves corresponds to the case of a static black hole without dark matter. Other curves in \ref{fig:O_DM} correspond to the different values of $\rho_0$ and in \ref{fig:O_Kerr} to the different values of $j$.}\label{fig:O_DM_Kerr}
\end{figure*}

In Fig.~\ref{fig:O_DM} the orbital angular velocity of test particles in the field of a static black hole without dark matter (solid curves) and in the field of a black hole surrounded by dark matter (dashed curves) for different values of $\rho_0^*=\rho_0/(10^{-5}AU^{-2})$ have been plotted as function of the radial distance. It can be seen that in the presence of dark matter, test particles on the accretion disk have smaller angular velocity in all range of the radial coordinate. Similar behaviour can be observed for test particles in the Kerr space time as shown Fig.~\ref{fig:O_Kerr}. Here $j$ is the dimensionless angular momentum (spin parameter) defined as $j=a/M$, where $a$ is the Kerr parameter and $M$ is the mass of the Kerr black hole. Our aim here is to compare the behaviour of a static black hole with dark matter with that of a rotating black hole in vacuum to asses the possibility that dark matter envelopes may mimic the effects of angular momentum. Therefore, we considered here only prograde (direct, co-rotating) orbits in the Kerr metric since they give smaller values of the ISCO radius $r_i$ (for details see \citep{1972ApJ...178..347B}), similarly to the static black hole immersed in dark matter. So, we see that for increasing $j$ the feature of $\Omega^*$ is analogous to that of the static space-time with dark matter and one can not distinguish the two space-times from Figs.\ref{fig:O_DM}-\ref{fig:O_Kerr}.

\begin{figure*}
\centering
\subfigure[Energy versus radial distance\label{fig:E_DM} in the static space-time]{\includegraphics[width=3.2in]{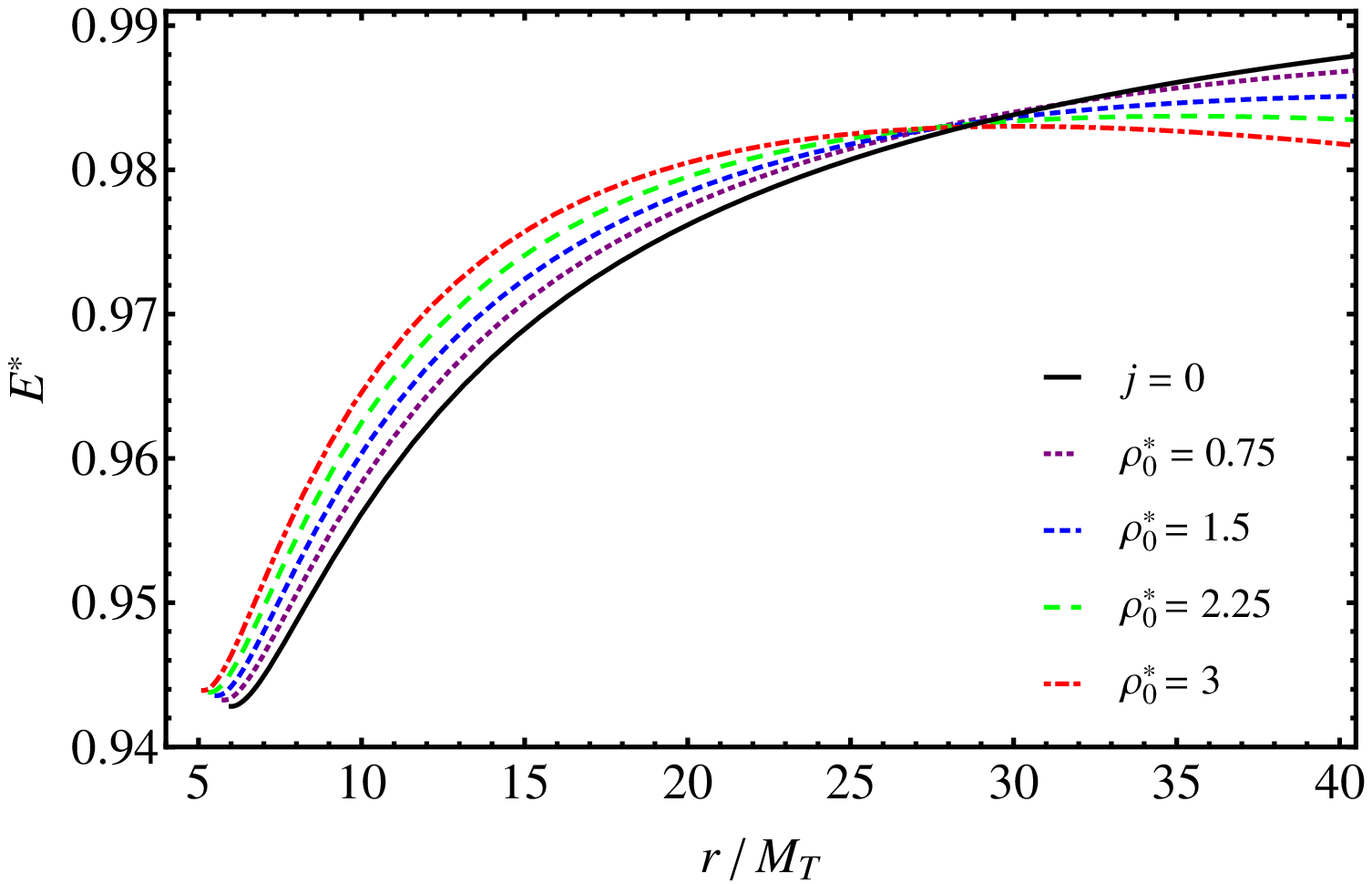}}
\subfigure[Energy versus radial distance \label{fig:E_Kerr} in the Kerr space-time]{\includegraphics[width=3.2in]{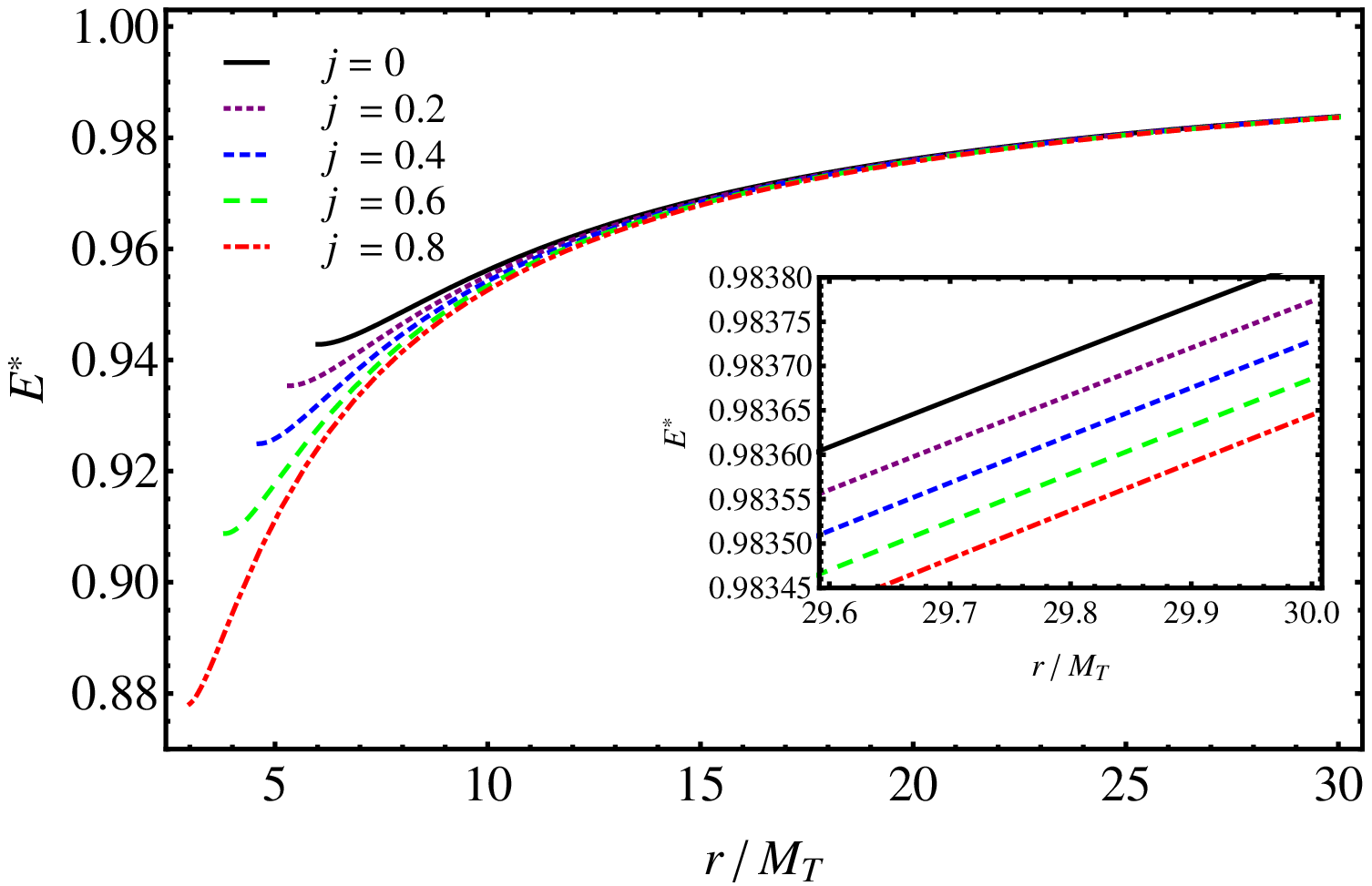}}
\caption{Numerical evaluation of energies $E^*$ in the static space-time with dark matter \ref{fig:E_DM} and in the Kerr space-time without dark matter \ref{fig:E_Kerr} of test particles in the accretion disk as functions of $r/M_T$. In both figures the solid curves show particles in the gravitational field of a static black hole without dark matter. Other curves in \ref{fig:E_DM} correspond to the different values of $\rho_0$ and in \ref{fig:E_Kerr} to the different values of $j$.}
\label{fig:E_DM_Kerr}
\end{figure*}

In Fig.~\ref{fig:E_DM_Kerr} the energy of test particles has been plotted as function of the radial distance. The legends are the same as in Fig.~\ref{fig:O_DM_Kerr}. It can be seen in Fig~\ref{fig:E_DM} that in the presence of dark matter, test particles on the accretion disk have larger energy at small radii and smaller energy at large radii as compared to particles on accretion disks around a static black hole $j=0$. However, in the Kerr space-time without dark matter for increasing $j$ the energy is always smaller than $j=0$ case. Therefore the two scenarios have distinguishable features in this case. Note, the energy has been calculated from the ISCO and to larger radii.

\begin{figure*}
\centering
\subfigure[Angular momentum versus radial distance\label{fig:L_DM} in the static space-time]{\includegraphics[width=3.2in]{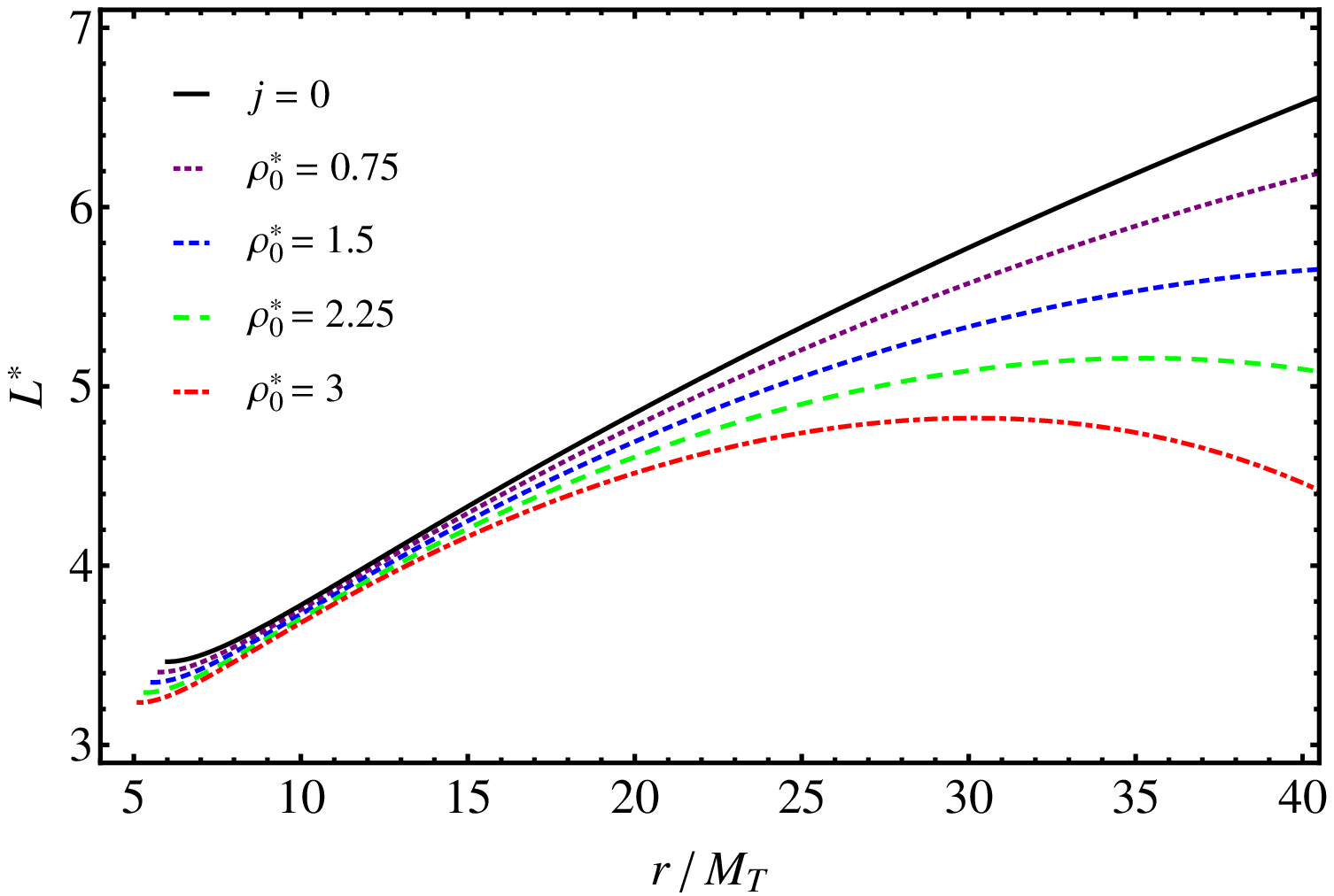}}
\subfigure[Angular momentum versus radial distance\label{fig:L_Kerr} in the Kerr space-time]{\includegraphics[width=3.2in]{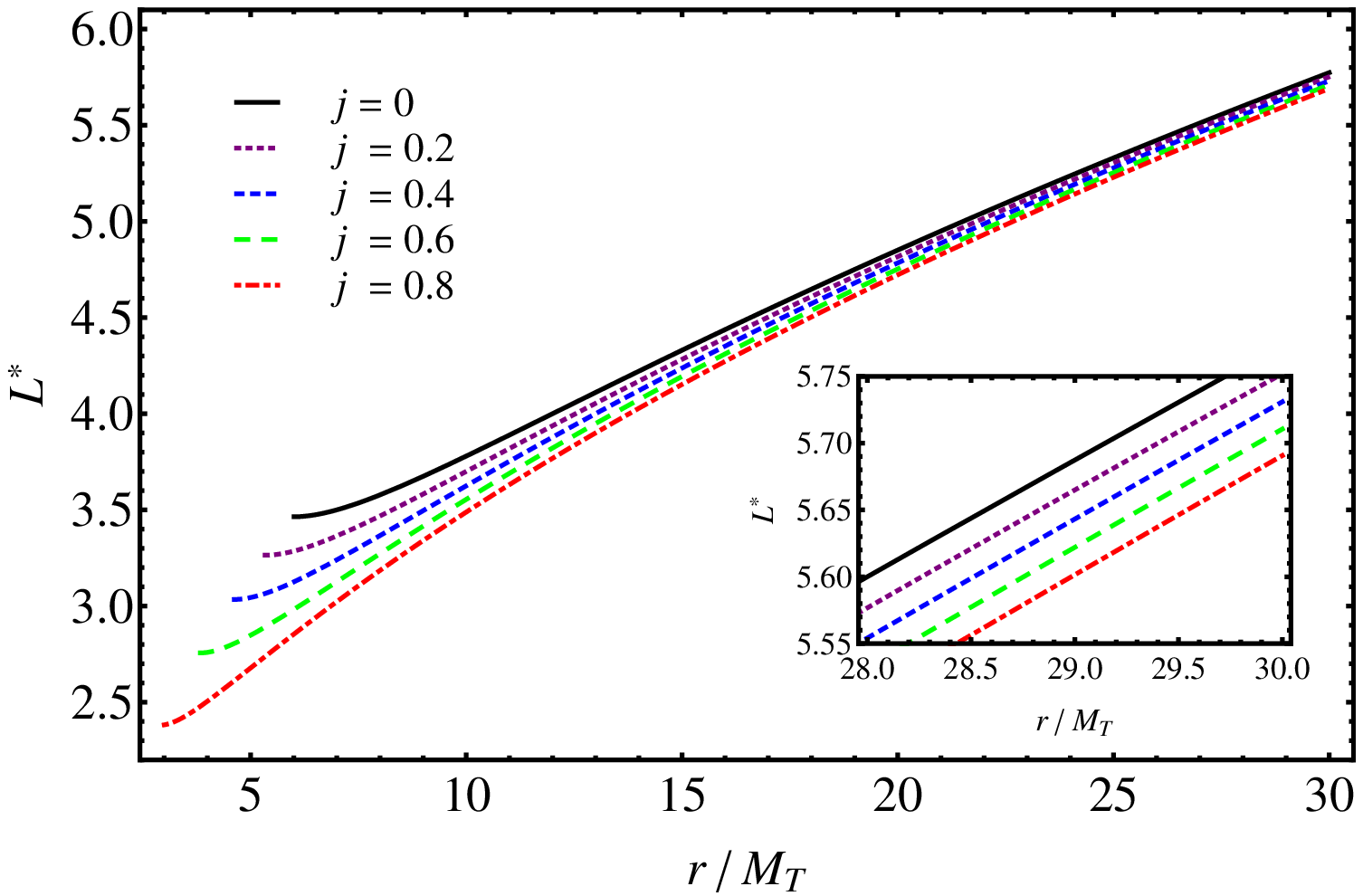}}
\caption{Numerical evaluation of orbital angular momentum $L^*$ of test particles in the accretion disk as functions of $r/M_T$. In both figures the solid curves show $L^*$ for particles in the gravitational field of a static black hole without dark matter. The remaining curves in \ref{fig:L_DM} correspond to the different values of $\rho_0$ and in \ref{fig:L_Kerr} to the different values of $j$.}
\label{fig:L_DM_Kerr}
\end{figure*}

In Figs.~\ref{fig:L_DM}-\ref{fig:L_Kerr} the orbital angular momentum of test particles have been plotted as function of the radial distance. The legends are the same as shown in Fig.~\ref{fig:O_DM_Kerr}. It can be seen that in the presence of dark matter, the angular momentum departs from the black hole case at large radii Figs.~\ref{fig:L_DM}, instead in the Kerr space-time the opposite occurs and the angular momentum converges to the static case as $r$ increases. In both cases $L^{*}$ is larger for a static black hole case.

\begin{figure*}
\centering
\subfigure[Radiating flux versus radial distance \label{fig:F_DM} in the static space-time]{\includegraphics[width=3.2in]{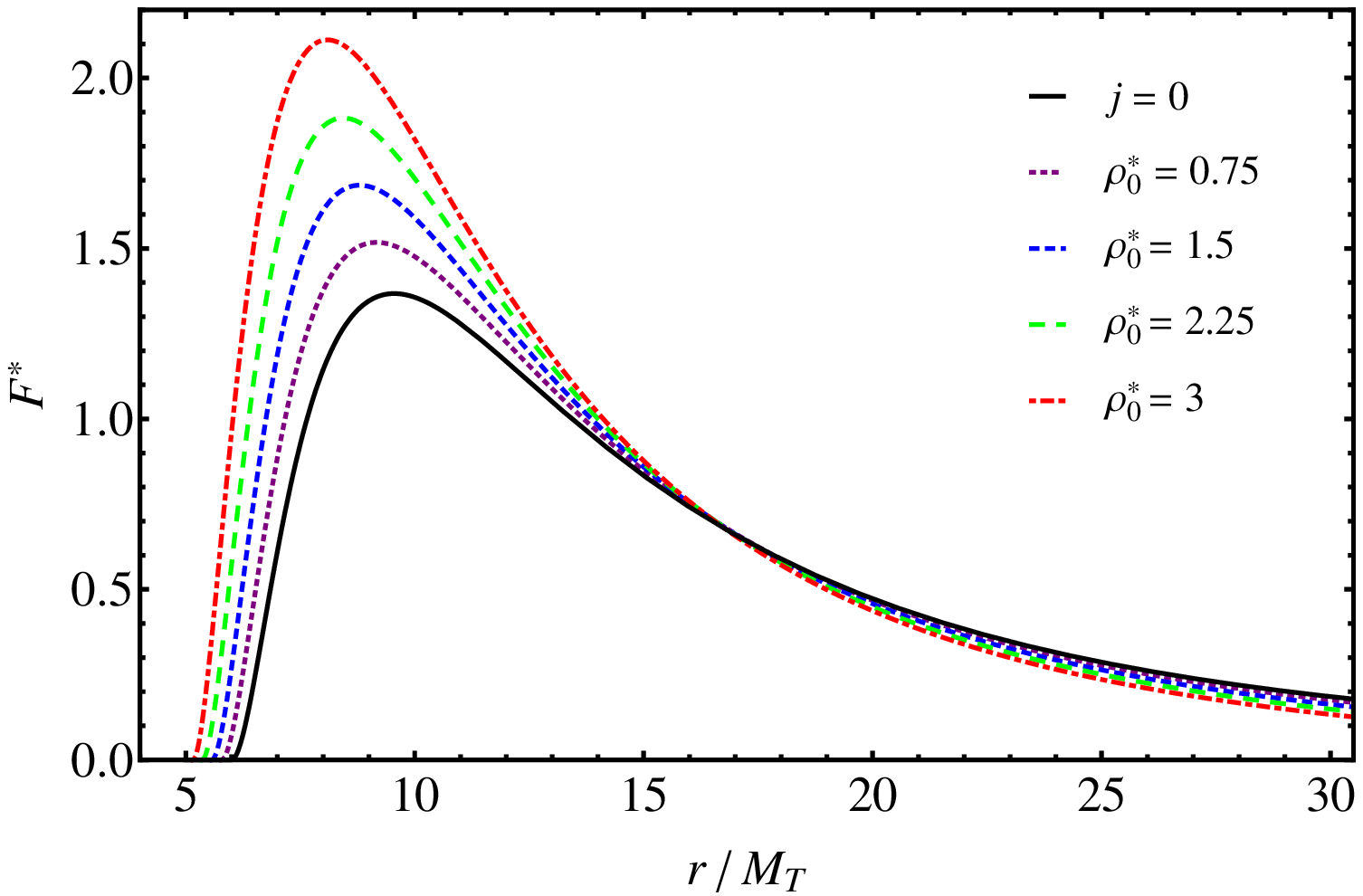}}
\subfigure[Radiating flux versus radial distance\label{fig:F_Kerr} in the Kerr space-time]{\includegraphics[width=3.2in]{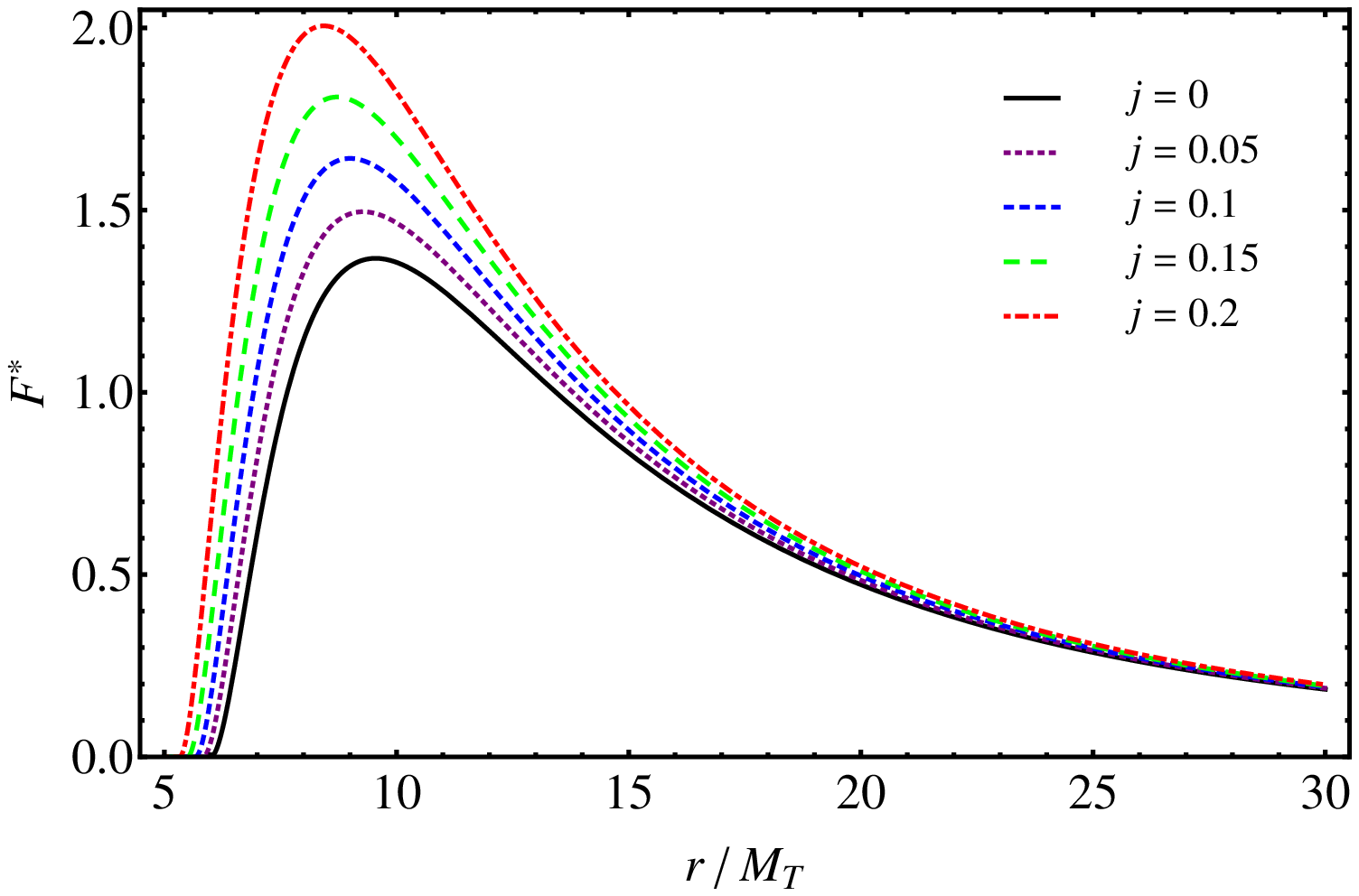}}
\caption{Numerical evaluation of the flux ${\cal F}$ divided by $10^{-5}$ of the accretion disk as functions of $r/M_T$. The solid curves in both panels correspond to the case of Schwarzschild in vacuum. The remaining curves in \ref{fig:F_DM} correspond to the different values of $\rho_0$ and in \ref{fig:F_Kerr} to the different values of $j$.}
\label{fig:F_DM_Kerr}
\end{figure*}

In Fig.~\ref{fig:F_DM_Kerr} the flux produced by the accretion disk is evaluated as a function of the dimensionless radial distance. Due to the fact that for dark matter $r_i<6M_{BH}$ the flux is higher with respect to the black hole without dark matter as the test particles, being able to stay on circular orbits at smaller radii, will emit more radiation before falling towards the central object (Fig.~\ref{fig:F_DM}).
Notice that since ${\cal F}(r_i)=0$ the value of the innermost stable circular orbit determines also the smallest radial distance from which particle emission can occur. Similar situation can be observed in the Kerr space-time without dark matter with co-rotating disk, which allows circular orbits to remain at smaller radii (Fig.~\ref{fig:F_Kerr}). However at larger radii the two scenarios can in principle be distinguished.

\begin{figure*}
\centering
\subfigure[Differential luminosity versus radial distance \label{fig:diff_Lum_DM} in the static space-time]{\includegraphics[width=3.2in]{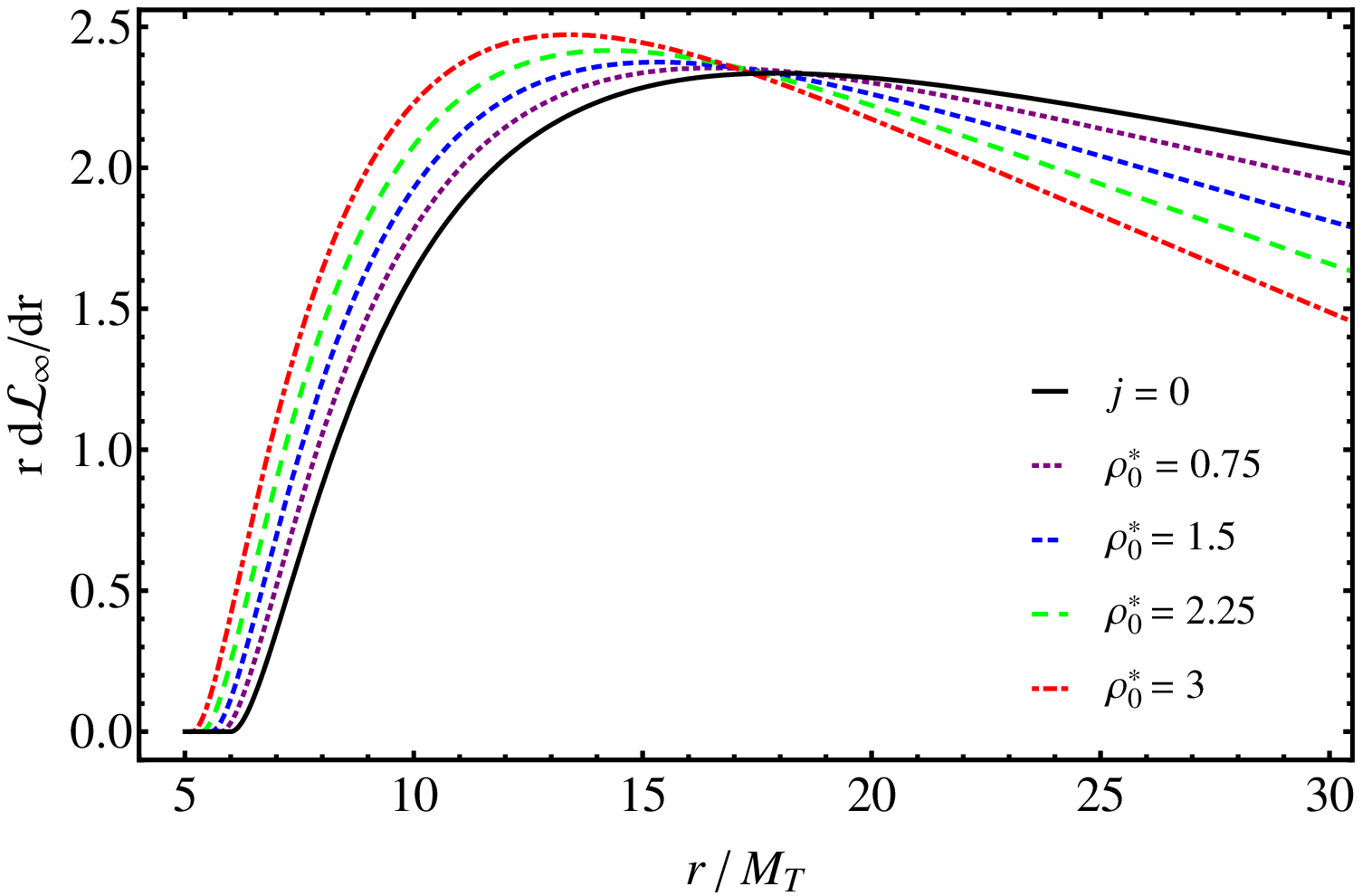}}
\subfigure[Differential luminosity versus radial distance in the Kerr space-time \label{fig:diff_Lum_Kerr}]{\includegraphics[width=3.2in]{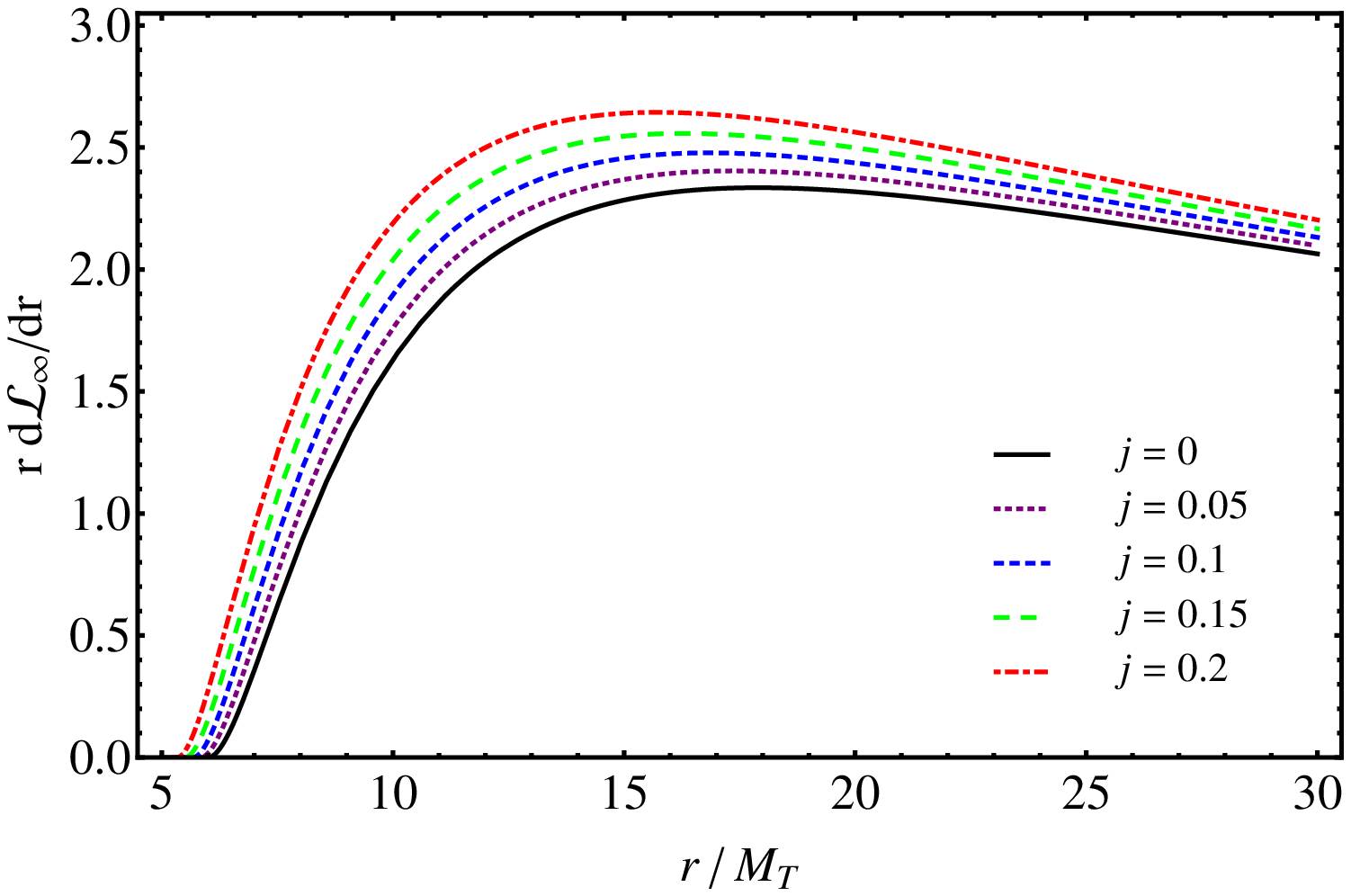}}
\caption{Numerical evaluation of the differential luminosity scaled in powers of $10^{-2}$ as a function of $r/M_T$. The solid curves in both panels correspond to the case of Schwarzschild in vacuum. The remaining curves in \ref{fig:diff_Lum_DM} correspond to the different values of $\rho_0$ and in \ref{fig:diff_Lum_Kerr} to the different values of $j$.}
\label{fig:diff_Lum_DM_Kerr}
\end{figure*}

Of course, when making observations of astrophysical sources, one does not measure directly any of the above quantities, and therefore they can not be used to distinguish the two scenarios in practical cases. On the other hand, measurements related to the disk's luminosity are possible.
In Fig.~\ref{fig:diff_Lum_DM_Kerr} the differential luminosity of the accretion disk is evaluated as a function of the dimensionless radial distance. Here we have a similar behavior as in Fig.~\ref{fig:F_DM_Kerr} at smaller radii owing to the fact that the ISCO is $r_i<6M_{BH}$ for both the static geometry with dark matter as well as for the Kerr geometry with positive values of $j>0$. However at larger radii the two space-times display different feature from each other.

\begin{figure*}
\centering
\subfigure[Spectral luminosity versus frequency of radiation in the static space-time with dark matter \label{fig:spec_Lum_DM}]{\includegraphics[width=3.2in]{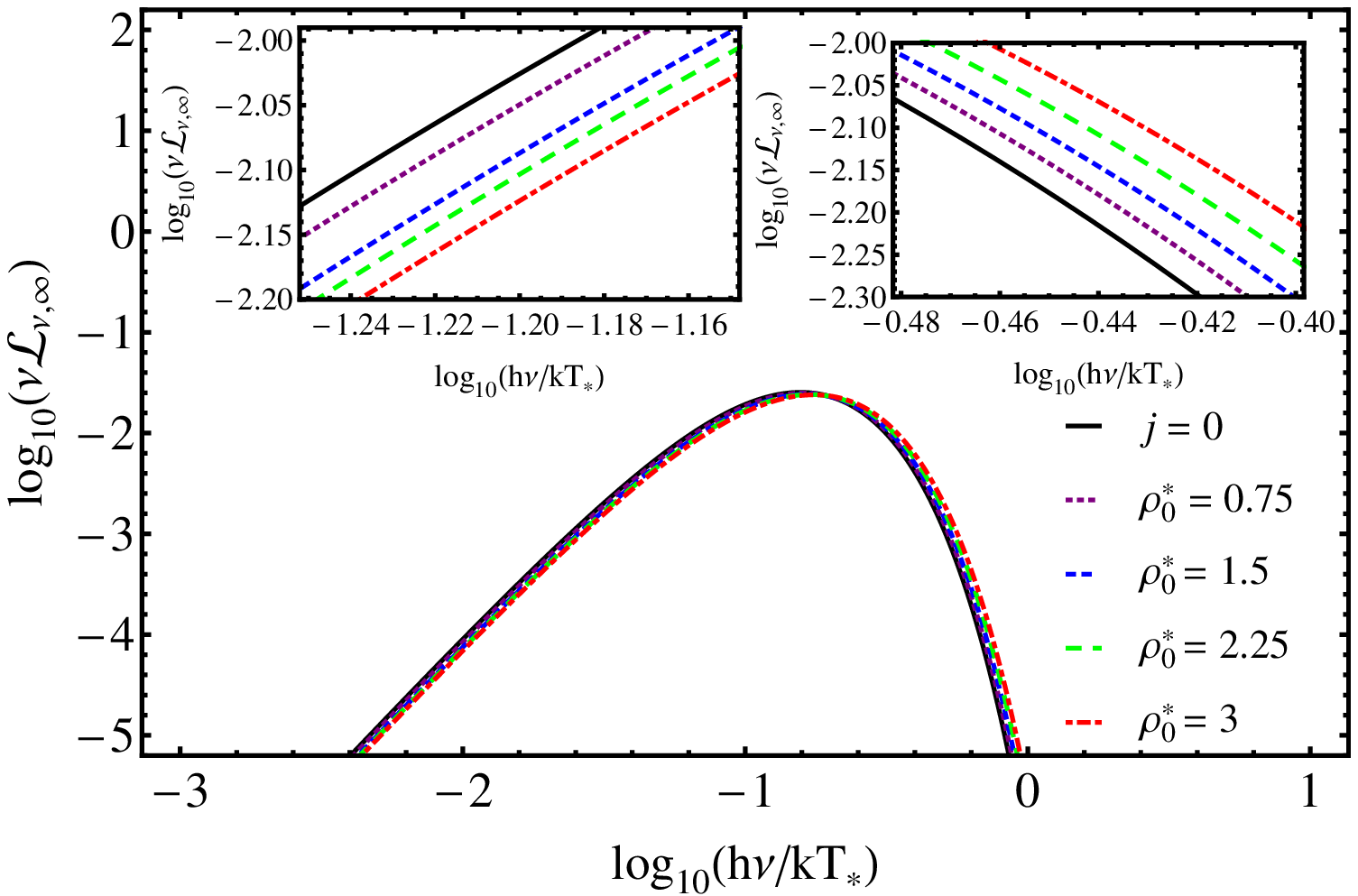}}
\subfigure[Spectral luminosity versus frequency of radiation\label{fig:spec_Lum_Kerr} in the Kerr space-time without dark matter]{\includegraphics[width=3.2in]{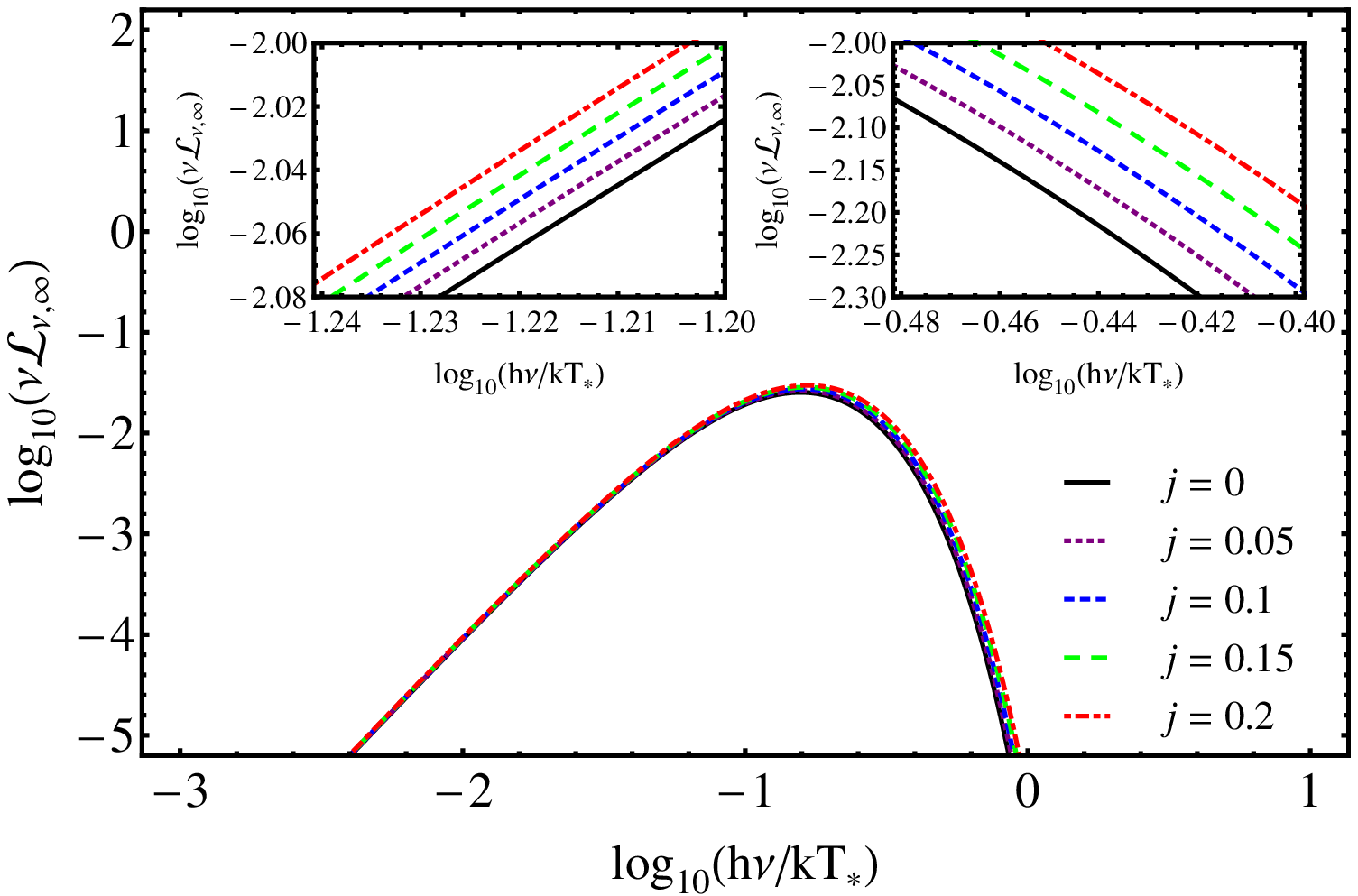}}
\caption{Numerical evaluation of the spectral luminosity of the accretion disk as a function of $h \nu /kT$ , i.e. as a function of frequency. In both figures the solid curves represents the case of a static black hole without dark matter. The remaining curves in \ref{fig:spec_Lum_DM} correspond to the different values of $\rho_0$ and in \ref{fig:spec_Lum_Kerr} to the different values of $j$.}
\label{fig:spec_Lum_DM_Kerr}
\end{figure*}

In Fig.~\ref{fig:spec_Lum_DM_Kerr} the spectral luminosity of the accretion disk is given as a function of radiation frequency $\nu$.

Here one can see the differences in the observable spectrum when dark matter is present (Fig.~\ref{fig:spec_Lum_DM}). One notable effect is a decrease in the spectral luminosity at low frequencies with respect to the case of a black hole in vacuum. At the same time, the most important effect comes from the fact that $r_i<6M_{BH}$ induces a larger spectral luminosity at high frequencies. This effect is due to the fact that particles in the inner regions of the accretion disk emit more energetic radiation at higher frequencies.

In the case of the Kerr space-time without dark matter for positive $j>0$ the spectral luminosity is always larger than for $j=0$ case. Indeed, this characteristics of the spectral luminosity may be used to test our model against the Kerr space-time at larger radii for smaller frequencies.

It is indeed interesting to investigate whether the dark matter envelope surrounding a static black hole can mimic the observable features of a rotating black hole in vacuum. In fact this turns out to be the case for a Kerr black hole with prograde accretion disk.

However, it should be noted that in principle the two geometries can be distinguished from the luminosity at large radii, since for Kerr $d{\cal L}_{\infty}/d\ln{r}$ approaches Schwarzschild from above for $j\rightarrow 0$, while in our model $d{\cal L}_{\infty}/d\ln{r}$ becomes smaller than the corresponding Schwarzschild value as $r$ grows.

\subsection{Gravitational lensing induced by dark matter}

An outstanding product of dark matter distribution, and another feature that may be used to test the model with observations, is its influence on gravitational lensing whereby the refraction index is modified due to the presence of dark matter. For example, it would be interesting to evaluate how much gravitational lensing of the back of the disk is expected for our dark matter distribution in comparison with the case of a black hole in vacuum. A detailed investigation on this topic goes beyond the main purposes of this work. However, here we can provide some arguments on the effect of dark matter on the amount of lensing by following the standard formalism of refraction index due to the dark matter gravitational field \citep{nclfp}. We assume that lensing is characterized by the superposition of the deflection angles of many infinitesimal point masses. Thus, invoking Fermat’s principle applied to the geodesic trajectories of our 4-dimensional curved space-time, we can find a  description of light rays in a gravitational field analogous to that of classical optics. In particular, for a transparent medium with a continuous refractive index, say $n$, where Fermat’s principle holds, the optical length is minimized by $\delta \int n(\tilde r)\left[d\tilde r^2+\tilde r^2d\tilde\Omega^2\right]=0 $, where $d\tilde\Omega^2=d \theta^2 + \sin^2 \theta d\varphi^2$ and $\tilde r$ is the isotropic coordinate in the equivalent optical metric, provided by
$ds^2=e^{2\Phi}\left(dt^2-n^2\left[d\tilde r^2+\tilde r^2d\tilde\Omega^2\right]\right)$. Comparing our metric, i.e. Eq. \eqref{eq:le}, with the isotropic coordinates above described,  we immediately find the system \citep{para}

\begin{equation}\label{eq:massprof3}
    \left\{
                \begin{array}{lll}
                  \frac{d\tilde r}{dr}=\frac{\tilde r}{r}e^{\frac{\Lambda}{2}},\\
                  \,\\
                n=\frac{r}{\tilde r}e^{-\frac{N}{2}}\,,
                \end{array}
              \right.
\end{equation}
and so the exact solution for $\tilde r$ in terms of $r$ is\footnote{Frequently this solution is approximated at large radii, where $\tilde r\simeq r$. In our case, however, we are interested in dark matter's envelope in the proximity of a black hole, so the approximation $\tilde r\simeq r$ does not hold.}
\begin{equation}\label{rtildeerre}
\tilde r = \exp\left\{\int\frac{dr}{r}\exp\left(\frac{\Lambda}{2}\right)\right\}\,.
\end{equation}

\begin{figure}
\centering
\includegraphics[width=1.1\columnwidth,clip]{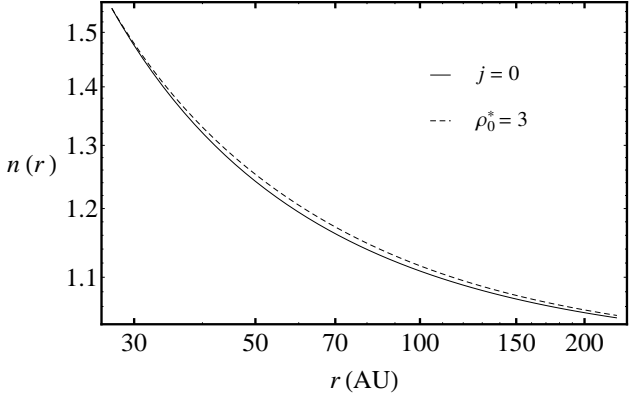}
\caption{Refractive index as a function of the radial coordinate.}\label{fig:refindex}
\end{figure}

\begin{figure}
\centering
\includegraphics[width=0.95\columnwidth,clip]{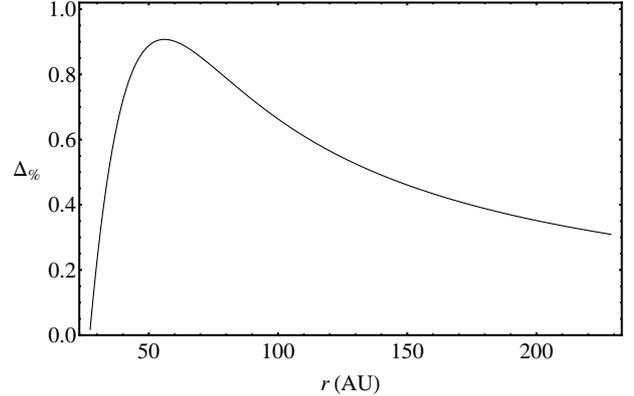}
\caption{$\Delta_{\%}$ as a function of the radial coordinate.}\label{fig:delta_r}
\end{figure}

\noindent In vacuum the spherical solution reduces to Schwarzschild, where $\Lambda\equiv-\ln\left(1-2M/r\right)$, with $M/r$ the standard Newtonian potential. It is therefore convenient to quantify the amount of dark matter lensing by means of the \emph{incremental function}:

\begin{equation}
\Delta_{\%}\equiv\frac{n_{DM}-n_{BH}}{n_{BH}}\,,
\end{equation}
where we have computed the second relation of the system \eqref{eq:massprof3}, substituting the relation \eqref{rtildeerre}. Thus, it is simple to notice that dark matter's amount influences the refractive directly through a correction related to both $\Lambda(r)$ and the gravitational potential of our model. The numerical behavior of $n(r)$ and $\Delta_{\%}(r)$ are portrayed in Figs.~\ref{fig:refindex} and \ref{fig:delta_r} respectively. In the dark matter envelope, the curve that includes dark matter  tends to $n\rightarrow1$ less strictly than the vacuum case. For the sake of clarity, however, the percentage evaluated by $\Delta_{\%}$ is very small and effectively $\leq1\%$, indicating that dark matter works as weak medium with small expected effects due to lensing.
By calculating $\Delta_{\%}(r)$ we practically exclude the black hole from the system and Fig.~\ref{fig:delta_r} essentially shows the potential of the dark matter envelope. Unlike a black hole, dark matter is distributed in a volume between $r_b$ and $r_s$, where $M_{DM}(r_b)=0$. As the radial distance $r$ increases the dark matter mass $M_{DM}(r)$ first grows faster, then grows slowly (see Fig. 3, right panel in \citep{bosh2019}). Therefore,  we observe maximum in Fig.~\ref{fig:delta_r} as the gravitational potential of dark matter is  $\sim{M_{DM}(r)/r}$.

\section{Conclusions}\label{sec:concl}

We considered a black hole surrounded by dark matter and we solved the TOV equations for the dark matter distribution extending for a finite thickness outside the black hole. We thus found allowed equilibrium configurations for dark matter with the inner boundary located either below or above the limit of the black hole's innermost stable circular orbit at $6M_{BH}$. The dark matter distribution in our toy model is static, even though it is understood that in more realistic model the dark matter particles should be dynamic and may fall onto the black hole once crossing the limit of bound orbits. However, since we are interested in accretion disks around super-massive black hole candidates we assumed that the timescales for all the dynamical processes involving the dark matter cloud's macroscopic quantities (such as density and inner boundary) are long as compared to the timescales for the processes occurring in the accretion disk, thus making the assumption of staticity acceptable.

With the aim to evaluate the role played by the presence of dark matter on the radiative flux, differential and spectral luminosity of the accretion disk we analysed the motion of test particles in the geometry produced by the black hole plus dark matter distribution.
We obtained the orbital parameters, such as angular velocity, energy and angular momentum per unit mass for test particles in the accretion disk.
We then evaluated the innermost stable circular orbit $r_i$ for different dark matter density profiles and used it to estimate the radiation flux emitted by the accretion disk. This study allowed us to qualitatively estimate and understand dark matter's effects on accretion disks. In particular, we showed that the dark matter envelope's thickness, which can be adjusted by varying the value of pressure $P(r_b)$ for any fixed density $\rho(r_b)$, and the location of the innermost stable circular orbit affect the disk's luminosity.
We determined the radiative flux and differential luminosity of the accretion
disks in the presence of dark matter and compared them with the case in the
absence of dark matter. It turns out that owing to the fact $r_i<6M_{BH}$ the
flux and differential luminosity are higher in the presence of dark matter.

This effect is similar to what is obtained for a rotating black hole with co-rotating accretion disk. In both cases the disk's spectral luminosity at high frequencies is larger than the pure Schwarzschild case.
Thus, in the presence of dark matter the spectral luminosity is greater at higher frequencies and smaller at lower frequencies with respect to the black hole in vacuum.
However, at small frequencies the spectral luminosity exhibits different behavior in our model as compared to the Kerr, suggesting that it may be possible, at least in principle, to experimentally test the validity of the model.

In fact, we showed that the presence of dark matter affects the geometry, which in turn affects the luminosity of the accretion disk.
As a consequence of our analysis, we suggest that super-massive black hole candidates in the distant universe, if surrounded by a sufficiently dense dark matter envelope, may be less massive than currently estimated, as their measured luminosity may be due partially to the presence of dark matter.

Obviously, the approach depends upon the choice of the background space-time, i.e. on the particular metric for the black hole candidate, and the dark matter's density profile. In the present article, for the sake of clarity, we limited the analysis to static black holes and a simple exponential density profile. However, similar considerations should readily extend to rotating black holes and more realistic density profiles.
In the future we aim at extending the work to the case of rotating black holes (i.e. the Kerr metric), black hole mimickers (such as static axially symmetric vacuum solutions) and exotic compact objects. Further, we will also focus on the form of dark matter density to understand how different dark matter distributions can influence the results. Since the density of normal matter and dark matter was higher in the early universe it is indeed possible that the accretion disks that formed around super-massive black holes behaved somehow differently from the simple models of black holes in vacuum that work for the present universe. Then our analysis can help to bring out the role played by dark matter distributions on the appearance of such objects.

\section*{Acknowledgements}
We would like to thank the editors for considering our work and the anonymous referee for the useful comments which helped to improve the quality of the article. O.L. acknowledges the support of INFN, iniziativa specifica MoonLIGHT-2. The work was supported in part by Nazarbayev University Faculty Development Competitive Research Grant No. 090118FD5348, by the MES of the RK, target program IRN: BR05236454, and by the MES Program IRN: BR05236494 and Grants IRN: AP05135753 and IRN: AP05134454.

\bibliographystyle{mnras}

%

\appendix

\section{Boundary value for pressure}
\label{app:A}
The equation of hydrostatic equilibrium in Newtonian gravity is given by
\begin{equation}
\frac{d P(r)}{d r}=-\rho(r)\frac{M(r)}{r^2}.
\end{equation}
Unlike in general relativity, this equation has an analytic solution in the classical limit
\begin{equation}
  P(r)|_r^{\infty}=-\int_r^{\infty}\rho(\tilde{r})\frac{M(\tilde{r})}{\tilde{r}^2} d\tilde{r}.
\end{equation}
Assuming that at infinity the pressure vanishes and taking into account Eqs. \eqref{eq:den} and \eqref{eq:massprof} one obtains
\begin{eqnarray}\label{eq:pressNewt}
&&P(r)=8\pi r_0^2 \rho_0^2\bigg[-e^{-2x}\left(\frac{1}{4}+\frac{1}{x}\right)-{\rm Ei}(-2x)+\\ \nonumber
&&+\left\{\frac{M_{BH}}{8\pi r_0^3 \rho_0}+ e^{-x_b}\left(1+ x_b+\frac{x_b^2}{2}\right)\right\}\left(\frac{e^{-x}}{x}+{\rm Ei}(-x)\right)\bigg],
\end{eqnarray}
where ${\rm Ei}(x)$ is the exponential integral for real non zero values of $x=r/r_0$
defined as
\begin{equation}
{\rm Ei}(x)=-\int_{-x}^\infty\frac{e^{-t}}{t}dt.
\end{equation}
For vanishing black hole mass $M_{BH}\to 0$ and boundary $r_b\to0$ one recovers the expression given by \citet{bosh2019}.
Hence, the pressure at the boundary $P_b=P(r_b)$ is defined when $r=r_b$ in Eq.~\eqref{eq:pressNewt}. Since the value of $P_b$ is not known in advance in general relativity, its Newtonian counterpart has been used in our computations as a test value. For fixed $\rho_0$ or correspondingly $\rho(r_b)$, the value of $P_b$ can be varied up to $1.5P_b$ due to the general relativistic corrections that have not been taken into account in the Newtonian equation. By varying $P_b$ one can vary the width/thickness of the dark matter envelope, hence the dark matter mass.


\end{document}